\documentclass[twocolumn,showpacs,preprintnumbers,amsmath,amssymb]{revtex4}
\usepackage{graphicx}
\usepackage{dcolumn}
\usepackage{bm}

\begin{document}

\title{Supercell band calculations and correlation for high-$T_C$ copper oxide superconductors.}

\author{T. Jarlborg}

\affiliation{
DPMC, University of Geneva, 24 Quai Ernest-Ansermet, CH-1211 Geneva 4,
Switzerland
}


\begin{abstract}
First principle band calculations based on local versions of density functional theory (DFT), together
with results from nearly free-electron models, can describe many typical
but unusual properties of the high-$T_C$ copper oxides.  The methods
and a few of the most important results are reviewed. Some additional
calculations are presented and the problems with
the commonly used approximate versions of 
DFT for oxides are discussed
with a few ideas for corrections. It is concluded that rather modest corrections
to the approximate DFT, without particular assumptions about strong correlation,
can push the ground state towards anti-ferro magnetic (AFM) order. 
Spin fluctuations interacting with phonons
are crucial for the mechanism of superconductivity in this scenario.

\end{abstract}

\pacs{74.25.Jb,74.20.-z,74.20.Mn,74.72,-h}

\maketitle

\section{Introduction.}

It is often assumed that strong correlation is important for an
understanding of the high-T$_C$ problem \cite{saw,oren,dama}. The failure of 
the local approximations to the exchange-correlation functional, such as
the local (spin) density approximation, LDA or LSDA 
\cite{lda,lsda,gga,sla} to produce the anti-ferro magnetic (AFM) insulating state of
many undoped transition metal oxides is generally quoted as the essential
reason for discarding DFT calculations for high-T$_C$ copper oxides \cite{dama}.
DFT is essentially exact, but the approximations to make it practical for
use in real applications seem inappropriate for oxides.
However, despite this problem there are several LDA results that fit
to the observed high-T$_C$ properties, provided that doping and supercells are
considered in order to account for imperfect lattice conditions such
as stripes, phonons or spin waves \cite{tj1,tj3,tj5}. Here is presented a short review
of those DFT results together with a discussion of the correlation problem
and an attempt to include further corrections to LSDA. Estimations of couplings
$\lambda$ caused by spin-phonon coupling (SPC) and pure spin fluctuations
are also made.


\section{Nearly free-electron model.}

The spin polarized potential from an AFM order (like that of undoped La$_2$CuO$_4$)
can be generated from  $V_{AFM}(x) = V_0 exp(-i\vec{Q}\cdot\vec{x})$,
where $\vec{Q}= \pi/a_0$, is at the Brillouin Zone (BZ) boundary, so that
there is one Cu site with spin up and one with spin down within a
distance $a_0$ ($a_0$ is the lattice constant). The potential of the
other spin has a phase shift of $\pi$. The band dispersion from
a 1-dimensional (1D) nearly free-electron model (NFE) is obtained from
a 2x2 eigenvalue problem, and it has a gap of 2$V_0$
at the zone boundary. Suppose now that an additional potential modulation
$V_{q}(x) = V_q exp(i\vec{q}\cdot\vec{x})$ exists, and that $\mid \vec{q} \mid
\ll \mid \vec{Q} \mid$, so that its periodicity or "wavelength" is much
larger than 2$a_0$. The product of these two modulations can
be written $V(x) = V_q exp(-i(\vec{Q}-\vec{q})\cdot\vec{x})$, where the two
amplitudes are combined into one coefficient, $V_q$. Fig. 1 shows the
real space configurations along the {100}-direction.
\begin{figure}
\includegraphics[height=6.0cm,width=8.0cm]{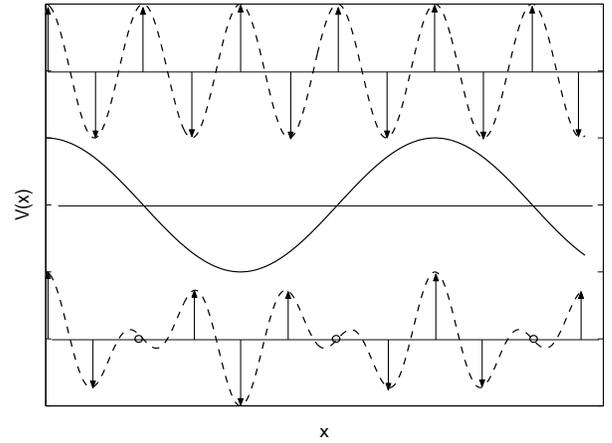}
\caption{A picture of how stripe like configurations are made up
from the product of two plane wave potentials. The arrows in the upper
row show the AFM moments on Cu sites given by the envelope
function $Re(exp(-i\vec{Q}\cdot\vec{x}))$ (broken line), and the unit 
cell contains 2 Cu sites along $\vec{x}$. The second row shows
the modulation $Re(exp(i\vec{q}\cdot\vec{x})$
(here $q=Q/4$) and the last row is the product
$Re(exp(-i(\vec{Q}-\vec{q})\cdot\vec{x})$ with the new
spin configurations on the Cu. The 
striped unit cell contains 8 Cu sites.
}
\label{fig0}
\end{figure}
 This potential
describes AFM modulations in stripes, with magnetic nodes (Cu with
zero moment) separated by $a_0Q/q$. The values of $V_q$ for
phonons or spin waves of different lengths are not known from the NFE
model, but some values will be fed in from the ab-initio Linear Muffin-Tin
Orbital (LMTO, \cite{lmto}) calculations, as will be described below. The LMTO
calculations are slow for large supercells. The cells need to be large
to cover the periodicity of realistic phonon and/or spin-waves, and
so far it has been possible to extend the cells in one direction only,
usually the CuO bond direction.  In contrast, the NFE model is very simple
and it can easily be extended to two dimensions (2D). The band dispersion
for k-points $(k_x,k_y)$ is now obtained from the eigenvalues of a 3x3 matrix \cite{tj6}.
Not only stripe order, but also "checkerboard" configuration can be modeled.

\begin{figure}
\includegraphics[height=6.0cm,width=8.0cm]{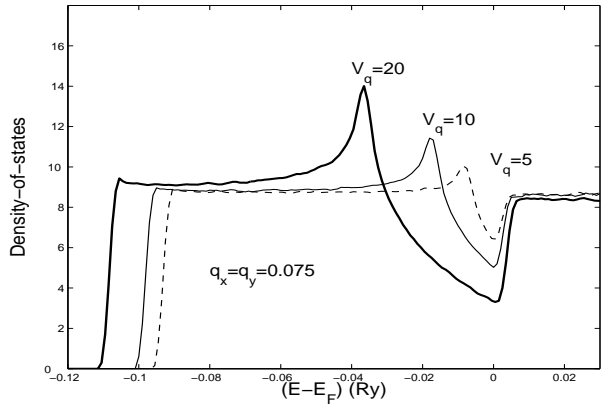}
\caption{Examples of the DOS from the 2D NFE model for 3 different strengths of
$V_q$ (in mRy) for $\vec{q}=(0.075,0.075)$. 
}
\label{fig1}
\end{figure}

Examples of the density-of-states (DOS) for 3 different $V_q$ are shown in
figure 2. The potential modulations are parallel to the CuO bond directions along $\vec{x}$
and $\vec{y}$, so the gaps appear on the $(k_x,0)$ and $(0,k_y)$ lines, as shown in figure 3. 
However, almost no effect from $V_q$ appears on the band dispersion in the diagonal
direction (along $(k,k)$). Therefore, the Fermi surface (FS) is not affected
along the diagonal, but fluctuations, described through amplitude variations
of $V_q$ for different positions and different time, will make the FS smeared
in the two bond directions. An example of this is shown in figure 4.

\section{Ab-initio 1D-LMTO.}

By 1D (1-dimensional) LMTO we mean ab-initio LMTO calculations 
for long (and narrow) supercells most often oriented
along the CuO bond direction \cite{tj2,tj8}. These calculations are based on the
local version of DFT, the local (spin) density approximation, LDA or LSDA \cite{lda,lsda}.
Cells with phonon distortions and/or spin waves within lengths of
4,8 and 12 lattice constants are typically considered in these calculations. 
The wavelengths of spin waves are
twice as long as those of phonons. Hole doping, $h$, in La$_{(2-h)}$Ba$_h$CuO$_4$
(LBCO) is generally modeled by the virtual crystal approximation (VCA)
where the nuclear and electronic La-charges (57.0) are reduced to (57-$h/2$) to account for
a perfectly delocalized doping ($h$ in holes per Cu). But some calculations
with real La/Ba substitutions show that improved superconducting properties
can be expected from periodic doping distributions \cite{apl}.
The maximal phonon distortion, $u_i$, for different sites, $i$, and AFM magnetic
moments on Cu, $m$, depend on temperature, $T$, force constants and spin stiffness.
Appropriate values of $u_i(T)$ and $m(T)$ at $T\approx 100$K are
deduced from experiments and calculations \cite{hum,thom,chen,coh,and,san}.

\begin{figure}
\includegraphics[height=6.0cm,width=8.0cm]{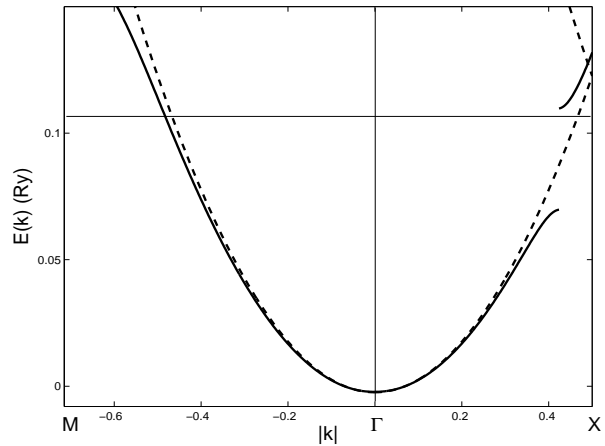}
\caption{An example of the 2D NFE bands along $(k,0)$ and $(k,k)$ for
$q_x = q_y = 0.075$ and $V_q=$ 20 mRy (full line). The FE band
without potential perturbations ($V_q=$0) is shown by the broken line. 
The thin horizontal line is at the DOS minimum, which is
situated at 0.15 holes/cell.  
}
\label{fig2}
\end{figure}

\begin{figure}
\includegraphics[height=6.3cm,width=8.3cm]{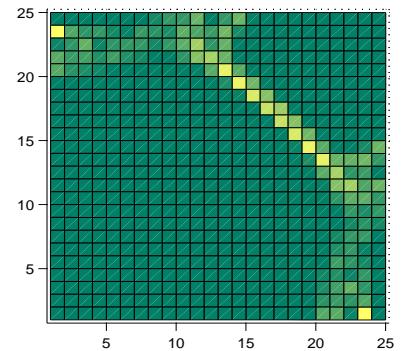}
\caption{An example of the Fermi surface in the 2D NFE model with fluctuations of $V_q$. 
The FS is well defined in the diagonal direction, while the fluctuations makes it
diffuse in the $x$ and $y$-directions.}
\label{fig3}
\end{figure}

A striking result of these calculations (also made for HgBa$_2$CuO$_4$) is that
a gap (or a pseudogap) will open up in the DOS, very similar to what
is found in the NFE models. The gap will appear near to $E_F$ for the
undoped material if the wavelengths ($\Lambda$) are long, while for short
phonons or spin waves the gap moves to lower energy \cite{tj1}. In summary,
$\Lambda = 1/h$, where $\Lambda$ is in units of $a_0$ and $h$ is the number
of holes per Cu.

A second important result is that the spin waves will be stronger, and the
pseudogaps will be deeper, if the waves coexist with phonon distortions of the
correct wave length and phase \cite{tj2}. Interactions between phonons
and spin waves have also been suggested in order to explain neutron scattering \cite{tj9,egami}.
Phonons like the "half-breathing" O-mode
and modes with z-displacements of La and apical oxygens are most effective
in this process of SPC. This mechanism offers
direct explanations of phonon softening, q-dependent spin excitations
and various isotope effects \cite{tj5,tj8}. 

The rather few LMTO results permit to establish only a few values of
$V_q$ for different $q$,$i$ of the phonons and spin waves. A general
trend of larger $V_q$ for long wave lengths seems clear, both for
phonons and for spin waves. A procedure based on the partial
character of the states above and below the gap in the NFE model
confirms this trend for spin waves, and it permits to fill in some
of the q-dependent points not calculated by LMTO. In all this gives
some confidence in the $q,i$-variations of $V_q$. However,
one can expect more vivid variations of the spin part near 
$q \rightarrow 0$, because LDA underestimates the transition towards
AFM. 

When these $V_q$-values are fed into the 2D NFE model it gives an
approximately linear variation of $q$ as function of $h$, up to a
saturation near $q \sim 0.125$ for $h$ larger than $\sim 0.13$ \cite{tj6}.
Hence, the linear dependence for low $h$ is in 
qualitative agreement with the result
from 1D-LMTO, but the pace is different. This is probably because
the gap opens $\it{both}$ along $x$ and $y$ in the 2D NFE model,
which makes the progression of the gap a bit slower than in 1D.
The saturation is because the $V_q$'s decrease for increasing
$q$ and it becomes impossible to open a gap at the low energy
where $E_F$ should be for large $h$. However, if $q_x$ and
$q_y$ are assumed to be different in the 2D NFE model it
is possible to follow one gap towards lower energy. A second
weaker gap moves to higher energy.

\section{Superconductivity}

The emerging picture from these band results is that moderately strong spin fluctuations,
which exist for many combinations of $\vec{q}$ and $\omega$, will be enhanced through
SPC to some particular phonon distortions. The selection
depends on the atomic character of the phonon mode as well as on doping and
wave length (i.e. on $\vec{q}$). The latter is because the pseudogap appears
at $E_F$ so that a maximum amount of kinetic energy can be gained from the
SPC mode. This is for low T when the states below the gap is of one spin
and well separated from the (unoccupied) state of the other spin above the gap.  
But for increasing T there will be mixed occupations of the states around
$E_F$ through the Fermi-Dirac function (and through thermal disorder), which will
decrease the spin density. This will decrease the spin polarization of the 
potential, which in turn will decrease the spin density even more, and at some
temperature T$^*$ the support of the pseudo gap from the spin wave will collapse \cite{tj5}.
The fact that $V_q$ is largest at low doping favors larger T$^*$ when $h \rightarrow 0$.
A high DOS at $E_F$ is important for a high superconducting $T_C$, and therefore is the
pseudogap in competition with  superconductivity in under-doped
cuprates \cite{tj6}. 

It was suggested that non-adiabatic electron-phonon coupling could be enhanced in the cuprates,
because of the low Fermi velocity in the z-direction \cite{noad}. This velocity
is comparable to the vibrational velocity and the electronic screening appeared
to be insufficient. However, the screening can be made by other electrons moving
within the planes. Moreover, phonon frequencies calculated within LDA
without assumptions of incomplete screening, agree satisfactory with
experimental frequencies \cite{chen,coh,and}. Instead,
excitations of virtual phonons coupled to spin waves can be important
for the  mechanism of superconductivity, but 
SPC makes the separation between pure electron-phonon coupling, $\lambda_{ep}$, and 
$\lambda$ caused by spin fluctuations, $\lambda_{sf}$,
less clear. 

The important observation is that atomic phonon distortions will trigger enhancement
of spin waves. If so, the possibility for larger $\lambda_{ep}$ is open, because
instead for the common approach to ignore spin effects in electron-phonon coupling,
there are larger matrix elements when the spin polarized part of the potential
is involved. This is readily imagined for a system which is
nonmagnetic when phonon distortions are absent, but magnetic when the distortions are
present. An estimation of pure $\lambda_{ep}$ and the coupling parameter for SPC, 
$\lambda_{SPC}$, in LBCO has been
presented earlier \cite{tj9}. With a total $N(E_F) \approx 0.9 (eV \cdot cell \cdot spin)^{-1}$,
and distortion amplitudes and potential shifts as in ref. \cite{tj6} this leads to $\lambda's$ of
the order 0.36 and 0.6 for pure phonons and SPC, respectively \cite{tj9}. These values appear
sufficiently large for a large $T_C$ in simple BCS-type formulations, but the precise
relation for $T_C$ depends also on other parameters \cite{aban}. 

\begin{table}[h]
\caption{\label{table1} 
Total energy ($\Delta E$, mRy per Cu), matrix element for spin fluctuations ($\Delta V$ in mRy), 
$\lambda_{sf}$ and local Stoner enhancement, $S$,
calculated for lattices of LBCO with two phonon distortions and without distortions.
}
\vskip 5mm
\begin{center}
\begin{tabular}{l c c c c c}
\hline
 phonon  & $\Delta E$ & $\Delta V$ & $\lambda_{sf}$ & $S$  \\
\hline \hline
 
 no phonon & 12.7 & 9 & 0.03 & 1.6  \\
 plane-O & 8.2 & 12.5 & 0.12 & 2.6   \\
 La & 7.4 & 12.0 & 0.11 & 2.4  \\

\hline
\end{tabular}
\end{center}
\end{table}

The third mechanism is that spin fluctuations work without coupling to
phonon excitations. Still, phonon distortions might be present, and will in that case
be important for enhancements of rapid (high energy) spin fluctuations. 
For instance, it was found that the local exchange enhancement (for AFM) on Cu sites are
largest when the surrounding atoms (La or oxygens) have been pushed away
from the Cu, and this leads to SPC between phonons and spin waves with equal
$\vec{q}$ and $\omega$ \cite{tj6}. But strong high-$\omega$ spin excitations should also be
possible on rows of Cu with enhanced exchange, perpendicular to the phonon.
Ab-initio calculations are not easy for such configurations because of the
large size of the required unit cells, but one can use some of the existing
results for first estimations of $\lambda_{sf}$ for rapid spin fluctuation
(so rapid that the phonon distortions appear to be static compared to
the spin fluctuation). 
The
enhancements depend on the different type of distortions as was discussed above. 
In this case we calculate $\lambda_{sf} = N \langle dV/dm \rangle^2 /(d^2E/dm^2)$, where $V$ is the potential,
$E$ is the total energy of the spin wave and $m$ is the magnetic moment (per Cu) \cite{tjfe}.
The difference in free energy between non-polarized and an induced (by magnetic field)
AFM wave can be written $E_m = E_0 + \kappa m^2$, so that $d^2E/dm^2 = 2\kappa$, which
permits to calculate $\lambda_{sf} = N \cdot (\Delta V)^2/\kappa$ for "harmonic"
spin fluctuations.
With parameters from the band results this makes $\lambda_{sf}$ equal to 0.03
when no lattice distortions are present, and 0.12 and 0.11 when the lattice
contain distortions, see Table 1.

The frequency has not been calculated explicitly, but a shortcut via exchange enhancements
can be used for a simple estimation,  as for FM fluctuations \cite{maz,tjfe}. 
Thus, $\omega_{sf} \approx 1/(4NS)$, where $S = \Delta V/ \Delta H$ is the local Stoner enhancement (on Cu)
and $\Delta H$ is the external field (~5 mRy in these calculations).
This makes $\omega_{sf}$ 100-200 meV depending on the type of lattice distortion.
Fig. 5 shows the combined results for direct $\lambda_{SPC}$ and indirect $\lambda_{sf}$ 
as function of frequency. The general shape is similar as the electron-boson
coupling function that has been extracted from recent optical spectra on Hg- and Bi-based
copper oxides \cite{heum}.

\begin{figure}
\includegraphics[height=6.3cm,width=8.3cm]{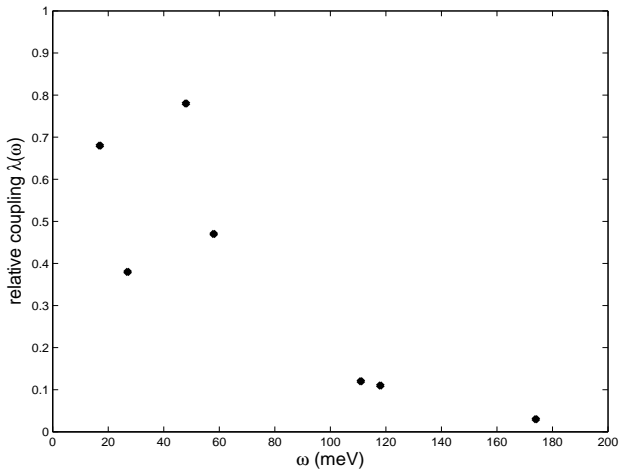}
\caption{The relative strength of $\lambda_{SPC}$ (the first
4 points) and $\lambda_{sf}$ (the 3 points at the highest energies).
The latter will move towards lower energy if the exchange enhancement
increases, while the points for SPC are fixed near the phonon
energies, see the text.}
\label{fig5}
\end{figure}

 The amplitudes of $\lambda_{SPC}$ and $\lambda_{sf}$ increase further 
if both  of the latter
distortions (plane oxygens {\it and} La) are considered in the calculation. 
The matrix element increases to about 14.5 mRy.
Despite this relative increase it is seen that pure spin fluctuations
give not as large $\lambda's$ as for SPC. This is especially clear when all lattice
modes are considered. However, these pure spin fluctuations can be
important for superconductivity, since they appear at large $\omega$.
In contrast, SPC is essentially limited to below the highest phonon frequency, i.e.
to 50-70 meV.
The relatively small absolute values of $\lambda_{sf}$
are partly caused by the use of LDA. Improved DF schemes which could predict
an AFM insulating state for the undoped systems, would lead to larger
exchange enhancements, larger $\lambda_{sf}$ and lower $\omega_{sf}$ in doped systems.

\section{Correlation and corrections to LDA.}

The question is if the real electron-electron correlation in the oxides is much larger
than what is included in LDA. The latter includes exchange and correlation (XC) for an electron gas
of varying density with the electron gas radius $r_s = 0.62 \rho^{-1/3}$ as parameter.
The density $\rho$ contains one electron within $r_s$. The largest values of $r_s$ 
($\approx 2 a.u.$) are
found in the low density valence-electron region between the atoms, but they are still
smaller than any typical atomic radius, $R_A$ in transition metals, their oxides and these
cuprates. This gives a major argument against using a strong on-site correlation parameter
to an atom, because the effect of an additional electron will readily be screened out, and is not 
noticed beyond $r_s$. In low-density materials however, such as the alkali metals,
$r_s \approx R_A$ and on-site correlation could be advocated. But LDA seems to
describe alkali metals well, which provides an additional argument in favor of local approximations of DFT.

The fact is that LSDA does not produce an AFM insulating ground state of several
oxides, so the problem is real and it may be related to the spin-polarized part
of the potential. Small errors with important consequences of this type are known
for LSDA. A well-known example is for Fe, where LSDA predicts nonmagnetic (NM)
face-centered cubic (fcc) lattice
as the ground state, whereas the generalized gradient approximation (GGA, \cite{gga}) correctly
finds the ferro magnetic (FM) body-centered cubic (bcc) ground state \cite{iron}. 
The difference between LDA and GGA is
not large concerning the bands, magnetic moments and other properties. 
The band properties of bcc Fe are quite good in calculations using both these
DFT functionals. But the order of the total energies for FM-bcc and NM-fcc is reversed
in the two cases.

An additional indication that the AFM insulating state of the cuprates is not far away
in LSDA is the fact that the AFM moment depends on the number of k-points. A sufficient number of k-points 
is required for convergence, but early works noticed that the AFM ground state became
stable if a coarse k-point mesh was used \cite{kpts,pick}. A similar conclusion
for the appearance of weak ferromagnetism in highly doped LBCO can be understood
from DOS effects and the Stoner criterion \cite{bba}. Thus, as for Fe, a small detail
can have large consequences.
It might be
that the total energy difference between two very different ground states (fcc nonmagnetic
and bcc-magnetic for Fe, and NM metallic and AFM insulating for cuprates) 
are very close, so that small errors in the computational details lead to the 'wrong' ground state.

Another indication of a nearby AFM state is that LMTO calculations with off-center linearization energies
can stabilize AFM in HgBa$_2$CuO$_4$ in calculations with sufficient number of
k-points. The normal procedure in LMTO is to chose linearization energies
at the center of the occupied sub bands, but if they are chosen at the bottom of
the bands, it leads to a slight localization of the states. An AFM order needs
less hybridization energy and AFM can be stabilized \cite{tj4}.
This is rather ad-hoc, since LMTO should not be made in such a way. A theoretical
justification is need for a better DFT scheme.

A first workable non-local density functional is the GGA
(generalized gradient correction \cite{gga}), where the gradient,
or the first derivative of the charge density, $d\rho(r)/dr$, brings out corrections
to the LSDA. The gradients are mainly radial in atoms and solids 
because of the steep drop
in charge density for increasing $r$ close to an atom. Only valence electrons have large
amplitudes in the interstitial region between the atoms, where the
gradients are moderate. However, higher gradients such as the second derivative
of the density can be large in the interstitial, a feature not
captured by the GGA. Nevertheless, by using GGA in the LMTO calculations for LBCO
there is some improvement, since the required critical magnetic field for
having a gap is reduced by about 25 percent compared to LSDA.

Densities with quadratic dependence as function of $r$ around a
fixed point at $r=0$ are used in a 2-particle model for calculations of 
correlation including corrections due to the
second derivatives. A non-interacting density with quadratic gradients
have the form
\begin{equation}
\rho(r)=\rho_0 + \frac{d^2\rho}{dr^2} r^2/2
\end{equation}
The Schr\"{o}dinger equation is solved for a density of electrons
surrounding a fixed electron at $r=0$, where it is taken into account that 
the effective mass is 1/2 \cite{corr,tjx}.
The interacting potential can be written
\begin{equation}
V(r)=e^2/r + \ell(\ell+1)/r^2 + \mu_{xc}(r) + V_{ext}(r)
\end{equation}
The exchange-correlation within the surrounding electron cloud
is taken into account through LDA ($\mu_{xc}$), and the external potential
$V_{ext}$ can include relative kinetic energy variations.
The two last terms in $V(r)$ are
not very important for the results for gradient variations. The strongest term is the
unscreened Coulomb repulsion that diverges at $r=0$. The second term, the centrifugal
term is included only for exchange (with $\ell=1$, higher
$\ell$ will have higher energy), when the Pauli principle requires
that the density for equal spin must be zero at $r=0$ \cite{tjx}. 

The solution $\Psi(E,r)$ of the Schr\"{o}dinger equation for this potential
is used to calculate the XC energy as 
the difference in Coulomb energy between interacting
and non-inteacting densities; $\mu = \int (\Psi(E,r)^2 - \rho(r))/r d^3r$.
The energy $E$ is determined from a required boundary condition of $\Psi(E,r_s)$.
The variation of the correlation
potential on the second gradient is quite easy to understand. If the gradient
is positive, so that there is a tendency for a 'hole' at $r$=0, it will
be less effective for the Coulomb repulsion to make the correlation hole
deeper at this point. Therefore, $\mu_c$ will decrease for positive density gradients
near the interstitial region. The gradient can change closer to the high density
regions near the nuclei and make the correlation larger than for a constant density, but
correlation becomes negligible in comparison to exchange when the density is high.
The C-correction is parametrized through the electron gas parameter $r_s$
and $Q=\rho(r_s)/\rho(0)$, so that
$C=1-5\sqrt{Q}/4+\sqrt{(r_s)}*Q/2$ for $Q > 1$
and as the inverse of this parametrization if $Q <1$. The expression is only
applied for $Q$ between -0.25 and 0.8, and $C-1$ is used as a scaling factor of the
correlation part from LDA. This correction makes the potential a little more
repulsive in the interstitial region. The Cu-d states becomes more localized,
which can promote AFM order, see Table II.

One fundamental theorem of DFT is that of "v-representability", that essentially says 
that there is a one-to-one correspondence between the density $\rho(r)$ and
potential $V(r)$ \cite{lda}. The only possibility is that two equivalent
charge densities can give an uninteresting shift in the potential, which
can be absorbed by the shift of the energy spectrum. Imagine now that
an external potential is applied to an electron gas of varying density $\rho(r)$.
The free electron wave functions will oscillate more if more kinetic energy
is given to the electron gas, or less if kinetic energy is subtracted, even if
$\rho$ remains constant. The exchange part of the potential (X)
will be modified because of the ability of the wave function to
readjust itself around a second electron of the same spin
(to create an exchange "hole"), as can
be imagined from Slater exchange \cite{sla} or the two-particle model \cite{tjx}. 
A similar change occurs for correlation (C), which is caused by the 
immediate Coulomb repulsion between all electron pairs. 
If $\rho$ is constant in space there 
is only an uninteresting potential shift, but for varying
charge densities there can be more subtle effects. The LDA is
derived from the principle of density variations at the Fermi
energy, and the relation between $E_F$ (which is a measure
of the kinetic energy) and density is 
given by;

\begin{equation}
E_F^{(1)}(r) = (3 \pi^2 \rho(r))^{2/3}
\end{equation}

This relation can be compared to $E_F^{(2)}(r)=E_F-V(r)$ within real atoms, where $E_F$
is from the band calculation with potential $V(r)$. Except
for $r$ very close to the nuclei there is not too large a difference
between $E_F^{(1)}(r)$ and $E_F^{(2)}(r)$.
In some regions the two values can be quite
different.  If so, it implies that the density is not in equilibrium with
the kinetic energy as was used in the derivation of the LDA potential. 
Oxygen sites in oxides are often negatively charged, which suggests
that the kinetic energy should be smaller than what comes out from
eq. 1. These effects are moderated by the fact that the charge transfer
is a result of an attractive potential. It turns out that positive
and negative deviations from eq. 1 exist in shells within all atoms.

Thus, band calculations with correction for the kinetic energy will
be attempted where the local exchange potential is written

\begin{equation}
\mu_{x}(f_e,r_s) = X(f_e,r_s) \mu_{KS}(r_s)
\end{equation}
where $X(f_e,r_s)$ is a scaling function of the normal Kohn-Sham potential $\mu_{KS}$,
and $f_e =  E_F^{(1)}(r_s)/E_F^{(2)}(r_s)$.

\begin{figure}
\includegraphics[height=6.3cm,width=8.3cm]{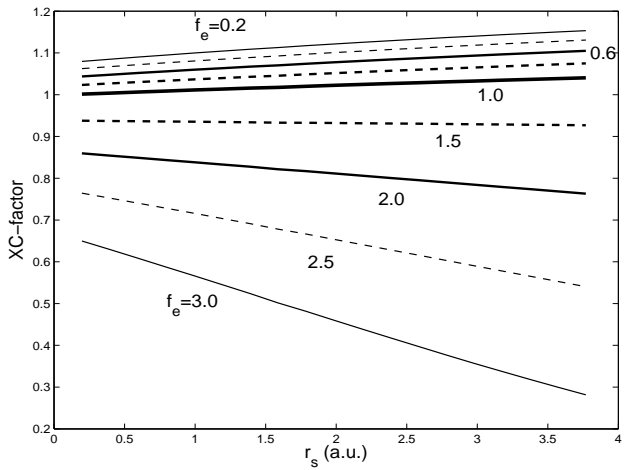}
\caption{The calculated values of the scaling function $X$ as function
of $r_s$ and $f_e$ (see text). 
}
\label{fig6}
\end{figure}

A derivation of the scaling function $X(f_e,r_s)$
can be done directly from the Slater functions
by use of energy shifts in the arguments of the plane waves \cite{sla}. However, negative
shifts leading to localized waves have to be avoided, and the resulting
$X$-function seems too sensitive to small variations of $f_e$. As
an alternative we determine the scaling function from a two-particle
model as in ref. \cite{tjx,corr}, with a renormalization to make $X(1,r_s)=1$.
The result is displayed in fig. 6 for the appropriate range of
$r_s$ and $f_e$. The real value of $f_e$ for $r \leq 0.05 R_{RWS}$ increases
and can be larger than 50 near the nuclei. This is extreme and $f_e$ is
here cut off at 3.

\begin{table}[h]
\caption{\label{table2} 
The required applied magnetic fields (in mRy on Cu) in order to obtain a
zero gap in LMTO calculations which
include the different corrections to the potential, see the text. 
}
\vskip 5mm
\begin{center}
\begin{tabular}{l c c c c c }
\hline
 LSDA  & GGA & GGA+C & GGA+X & GGA+C+X  \\
\hline \hline
 
 8.8 & 6.6 & 5.5 & 5.0 & 4.0   \\

\hline
\end{tabular}
\end{center}
\end{table}

In preliminary calculations for undoped LBCO we do a rescaling of the exchange
due to kinetic energy and of correlation due to second gradients. 
The comparison is made with standard LSDA
for an AFM unit cell where a staggered magnetic field is applied to
the Cu sites, and it is found that a field of $\pm$ 8.8 $mRy$ is
the limit for having a zero gap $(E_g \sim 0.25 mRy)$.
The magnetic moment is then $\pm 0.18 \mu_B$ per Cu.
The 
amplitude of the required magnetic
field to obtain a zero gap is reduced when the calculations include the
correction factors, see Table II.
The absolute values can depend on details of the band calculations \cite{tj4}, but
the trend towards a stability of an AFM state is clear. Corrections of this
type will be interesting for seeing enhanced spin fluctuations for long
wave length spin waves in doped cuprates. Further enhancements of $\lambda_{sf}$
can also be expected \cite{tj6}. Potential corrections must ultimately be
tested for other types of materials in order to verify that some
properties will not deteriorate. Here for the cuprates the corrections permit to
proceed in the modeling of more properties from the band results.

\section{Conclusion}

Results of band calculations for supercells with frozen phonons and spin waves 
suggest that a Fermi-liquid state can cause pseudo gaps and dynamic stripes. 
Together with a NFE model it is possible to describe the doping dependence
of many normal state properties of the cuprates. The coupling between
phonon distortions and spin fluctuations seems to be crucial for the mechanism
of superconductivity, so that the spin-polarized part of $\lambda$ is
most enhanced by simultaneous excitations of phonons and spin waves.
Two different mechanisms for superconductivity mediated by spin fluctuations are
possible. The largest coupling parameter is when a phonon is excited together
with the spin fluctuation. Lower couplings, $\lambda_{sf}$ at larger
energies, are independent of the phonon excitation, but these spin fluctuations 
can nevertheless profit from possible phonon distortions of the lattice.
Still, absolute numbers are too small when using LDA, as is also concluded
from the absence of AFM stability in LDA calculations for undoped systems.
However, it is argued that corrections due to higher order density gradients
and kinetic energy are able to bring the band calculations closer to AFM.
This is shown in calculations for undoped LBCO by yet very approximate 
corrections due to non locality and kinetic energy. Refinements of
such corrections will be of interest for application to supercells,
including doping, phonon distortions and spin waves, since it can be expected
that realistic $\lambda's$ will be obtained.




\end{document}